\documentclass[
reprint,
amsmath,amssymb,
pre,
floatfix,
]{revtex4-1}

\usepackage{graphicx}
\usepackage{dcolumn}
\usepackage{bm}

\usepackage[normalem]{ulem} 
\usepackage{titlesec}
\usepackage{dcolumn}
\usepackage{bm}
\usepackage{graphicx}
\usepackage{afterpage}
\usepackage{xcolor}
\usepackage{amsmath}
\usepackage{amssymb}
\usepackage{graphicx}
\usepackage{bm}
\usepackage{epstopdf}
\usepackage{hyperref}
\usepackage{csquotes}
\hypersetup{
	colorlinks=true,
	linkcolor=blue,
	filecolor=magenta,      
	urlcolor=blue,
	citecolor=red,
}

\newcommand{\Jav}{J_\mathrm{av}}

\begin{document}

\preprint{betacell1}

\title{The interplay between diversity and noise in an excitable cell network model}

\author{Stefano Scialla$^{(1,2)}$}\thanks{(corresponding author)}
\email[]{stefano.scialla@kbfi.ee}
\author{Marco Patriarca$^{(1)}$}
\email[]{marco.patriarca@kbfi.ee}
\author{Els Heinsalu$^{(1)}$}
\email[]{els.heinsalu@kbfi.ee}
\affiliation{{\rm (1)} National Institute of Chemical Physics and Biophysics -  Akadeemia tee 23, 12618 Tallinn, Estonia}
\affiliation{{\rm (2)} Department of Engineering, Universit\`a Campus Bio-Medico di Roma - Via \'A. del Portillo 21, 00128 Rome, Italy}

\vspace{0.5cm}

\date{\today}

\begin{abstract}
We study the interplay between diversity and noise in a 3D network of FitzHugh-Nagumo elements, with topology and dimensions chosen to model a pancreatic $\beta$-cell cluster, as an example of an excitable cell network. 
Our results show that diversity and noise are not equivalent sources of disorder but have different effects on network dynamics. 
Their synchronization mechanisms may act independently of one another or synergistically, depending on the mean value of the diversity distribution compared to the intrinsic oscillatory range of the network elements.
\end{abstract}

\maketitle

\section{Introduction}

The study of the beneficial role of disorder in a broad range of biological, physical, and chemical phenomena, has become a fundamental research topic in complex systems dynamics. 
A seminal work in the field was the introduction of stochastic resonance (SR) in the early eighties, initially proposed to explain the occurrence of Earth ice ages~\cite{Benzi-1981a} and later on studied by numerous authors across various disciplines~\cite{Nicolis-1981a,Nicolis-1982a,Benzi-1983a,McNamara-1988a,Gammaitoni-1989a,Matteucci-1989a,Jung-1991a,Gammaitoni-1991a,Mori-2002a,Rao-2002a}. SR can happen when a nonlinear system is driven simultaneously by a periodic external forcing and noise, resulting in an amplification of the system response to the external signal~\cite{Heinsalu-2009a}.

Some of the subsequent studies showed that significant noise-driven effects, analogous to SR, can be observed also without periodicity of the external signal~\cite{Gang-1993a}, and even in the absence of any external signal, as in self-induced stochastic resonance~\cite{MURATOV2005227,Marius-2017} and coherence resonance~\cite{Pikovsky-1997a}.
Coherence resonance (CR) is an ordered response of a nonlinear excitable system to an optimal noise amplitude, resulting in regular pulses. 
Beyond its effects on a single nonlinear unit, the role of noise was also extensively studied from the standpoint of its ability to improve synchronization in coupled oscillator networks, again both in the absence and in the presence of an external forcing~\cite{Jung-1995a,Liu-1999a,Busch-2003a}.
Broadly speaking, one may say that the first twenty years of research in this field focused, to a large degree, on the effects of noise in bistable or excitable systems~\cite{Gammaitoni-1998a,McDonnell-2009a}, comprising either several or just one element.

At the beginning of the new century, it was found that somewhat analogous effects to those of noise can be produced in networks of coupled oscillators through the heterogeneity of the oscillator population~\cite{Cartwright-2000a,Tessone-2006a}. 
This led to the introduction of diversity-induced resonance (DIR), which denotes the amplification of a network response to an external signal, driven by the heterogeneity of network elements~\cite{Tessone-2006a,Toral-2009a,Chen-2009a,Wu-2010a,Wu-2010b,Patriarca-2012a,Tessone-2013a,Grace-2014a,Patriarca-2015a,Liang-2020a}. Just like SR, also DIR can occur both in the presence and in the absence of an external forcing. 
In the latter case it has been named diversity-induced coherence~\cite{KAMAL-2015a}.

It should be clear from the above that SR and CR can occur even in systems made of a single element, therefore they are not intrinsically collective phenomena. Instead, by definition, DIR represents a collective disorder effect driven by population heterogeneity.

Most of the previous literature has emphasized either the analogies between SR and DIR~\cite{Tessone-2006a,Tessone-2007a}, considering them as two faces of the same medal, or the possibility to enhance resonance induced by noise thanks to diversity optimization (or vice versa)~\cite{degliesposti-2001a,Li-2012a,Li-2014a,Gassel-2007a}. 
Relatively little work~\cite{Zhou-2001a,Glatt-2008a} has been devoted to the implications of the above-mentioned intrinsic difference between the two phenomena, which has not yet been fully analyzed. In this paper, by systematically studying the prototypical model of a heterogeneous network of pancreatic $\beta$-cells~\cite{Cartwright-2000a,Scialla-2021a} with the addition of noise, we provide insights into the different mechanisms by which diversity and noise can have markedly distinct effects on network dynamics.

\section{Model}
\label{model}

As a paradigmatic example of a system of coupled nonlinear units, we investigate an excitable network of FitzHugh-Nagumo elements. 

Individual elements of such network are described by the dimensionless FitzHugh-Nagumo equations~\cite{Fitzhugh-1960a,FitzHugh-1961a,Nagumo-1962a,Cartwright-2000a,Scialla-2021a}:
\begin{eqnarray}
\dot{x} &=& a \left( x - x^3/3 + y \right) \, ,
	\label{eq_FN1a}
    \\
    \dot{y} &=& - \left( x  + by - J \right)/a .
	\label{eq_FN1b}
\end{eqnarray}
When modelling the behavior of a $\beta$-cell, the variable $x(t)$ represents the fast relaxing membrane potential, while $y(t)$ is a recovery variable mimicking the slow potassium channel gating. 
Depending on the value of $J$, the unit will be in an oscillatory state if $|J|< \varepsilon$ or in an excitable state if $|J| \ge \varepsilon$, where
\begin{equation}
\label{epsilon} 
\varepsilon = \frac{3 a^2 - 2 a^2 b -b^2}{3 a^3} \sqrt{a^2 - b} \, . 
\end{equation}
In addition to determining the width of the oscillatory interval $ (-\varepsilon,+\varepsilon)$, parameters $a$ and $b$ define the oscillation waveform and period.

Moving from the description of a single element to that of a heterogeneous 3D network of $N$ FitzHugh-Nagumo units, we assume a cubic lattice topology.
This implies that each element is coupled to its six nearest neighbors via a coupling term $C_{ij} (x_j - x_i)$, where $i$ and $j$ are indexes that identify an element $i$ and one of its coupled neighbors $j$, and $C_{ij}$ is the interaction strength. 
Notice that the choice of a cubic lattice topology is consistent with what is known about the architecture of $\beta$-cell clusters, where each cell is surrounded on average by 6-7 neighbor cells~\cite{PERSAUD-2014a,Nasteska-2018a}.
We make the simplifying assumption that the value of the coupling constants is the same for each network element, $C_{ij} \equiv C$ for any $i,j$. Since we want to study the interplay between diversity and noise, we also add a noise term $\xi_i(t)$ to the first equation.  
Then the corresponding FitzHugh-Nagumo equations for the $i$th element of the network are~\cite{Cartwright-2000a,Scialla-2021a}:
\begin{eqnarray}
\dot{x_i} &=& a \!\! \left[x_i \! - \! \frac{{x_i}^3}{3} \! + \! y_i + \! C\!\!  \sum_{j \in \{ n \}_i} \!\! (x_j \! - \! x_i) \! + \! \xi_i (t)\!  \right] \!\! , \label{FNEN1} \\
\dot{y_i} &=& -(x_i + by_i - J_i)/a \, . \label{FNEN2}
\end{eqnarray}
The sum over $j$ in Eq.~(\ref{FNEN1}) is limited to the set $\{n\}_i$ of the $n=6$ neighbors coupled to the $i$th oscillator.

The $J_i$ parameters in Eq.~(\ref{FNEN2}) are different for each network element and are used to introduce diversity; the $i$th element will be in an oscillatory state if $|J_i|< \varepsilon$ or in an excitable state if $|J_i| \ge \varepsilon$.
The $J_i$ values are drawn from a Gaussian distribution with standard deviation $\sigma_d$, mean value $\Jav$, 
and are randomly assigned to network elements. 
The standard deviation $\sigma_d$ will be used in what follows as a measure of oscillator population diversity, while the mean value $\Jav$ expresses how far the whole population is from the oscillatory range $ (-\varepsilon,+\varepsilon)$.

The term $\xi_i(t)$ in Eq.~(\ref{FNEN1}) is a Gaussian noise with zero mean, standard deviation $\sigma_n$, and correlation function $\langle \xi_i(t) \xi_j(t') \rangle = {\sigma_n}^2 \delta_{ij} \delta (t-t')$, meaning that $\xi_i(t)$ and $\xi_j(t)$ ($i \ne j$) are statistically independent of each other. The standard deviation $\sigma_n$ will be used in this work as a measure of noise applied to each network element.

The reason why we add $\xi_i(t)$ to the first equation for the fast variable is that this maximizes the effects of noise, making it easier to study its combination with diversity. 
Introducing $\xi_i(t)$ into the second equation would result in a minimal impact of noise on network synchronization~\cite{degliesposti-2001a}, due to the slower dynamics of the refractory variable. 
Noise effects would be mostly averaged out to zero by time integration and coupling, as we will show below with some numerical simulations.

The above model, which to our knowledge is studied here for the first time in the version we propose, can be used to mimic various excitable biological systems, such as pancreatic $\beta$-cell clusters and some types of neurons~\cite{Poznanski-1997a,Cartwright-2000a,Andreu-2000a,degliesposti-2001a,VRAGOVIC-2006a,Scialla-2021a}. 
In what follows we use the model to analyze the combined effect of diversity and noise, acting together on the same network. 
In particular, we are interested in potential synergies or antagonisms, as well as in a possible hierarchy between the two sources of disorder, which in spite of some analogies have fundamentally different synchronization mechanisms.

\section{Qualitative theoretical analysis}
\label{analysis}

The white noise $\xi_i(t)$ in Eq.~(\ref{FNEN1}) can represent a randomly fluctuating external current, which is able to shift the nullcline of the $x$ variable up or down and, therefore, to instantaneously change the position of the equilibrium point of each oscillator. 
Depending on the extent of the shift and on the value of $J_i$, this may result in a switch from a stable to an unstable equilibrium (or vice versa), corresponding to a transition from a resting to a spiking state of the oscillator (or vice versa).

This mechanism can be further illustrated, following Ref.~\cite{Tessone-2006a}, by introducing the global variables $X(t) = N^{-1} \sum_{i=1}^N x_i(t)$ and $Y(t) = N^{-1} \sum_{i=1}^N y_i(t)$.
We then define $\delta_i$ as the difference between $x_i$ and $X$, i.e., $x_i \equiv X + \delta_i$,
and introduce $M = N^{-1} \sum_{i=1}^N {\delta_i}^2$~\cite{Tessone-2006a,Desai-1978a}. 
Therefore, $M$ will increase when diversity increases. 
By averaging Eqs.~(\ref{FNEN1}) and (\ref{FNEN2}) over all $N$ network elements, 
we obtain the equations for the global variables $X$, $Y$:
\begin{eqnarray}
\dot{X} &=& a  \left[X(1-M)  -  X^3 /3 + Y +  \xi_G (t)  \right] , \label{FNENG1} \\
\dot{Y} &=& -(X + bY - \Jav)/a \, . \label{FNENG2}
\end{eqnarray}
Here noise effects are represented by a global white noise term $\xi_G(t) = N^{-1} \sum_i \xi_i(t)$ with zero mean
and correlation function $\langle \xi_G(t) \xi_G(t') \rangle = N^{-1} \sigma_n^2 \delta (t-t')$.

\begin{figure}[!t]
\includegraphics[width=8.8cm]{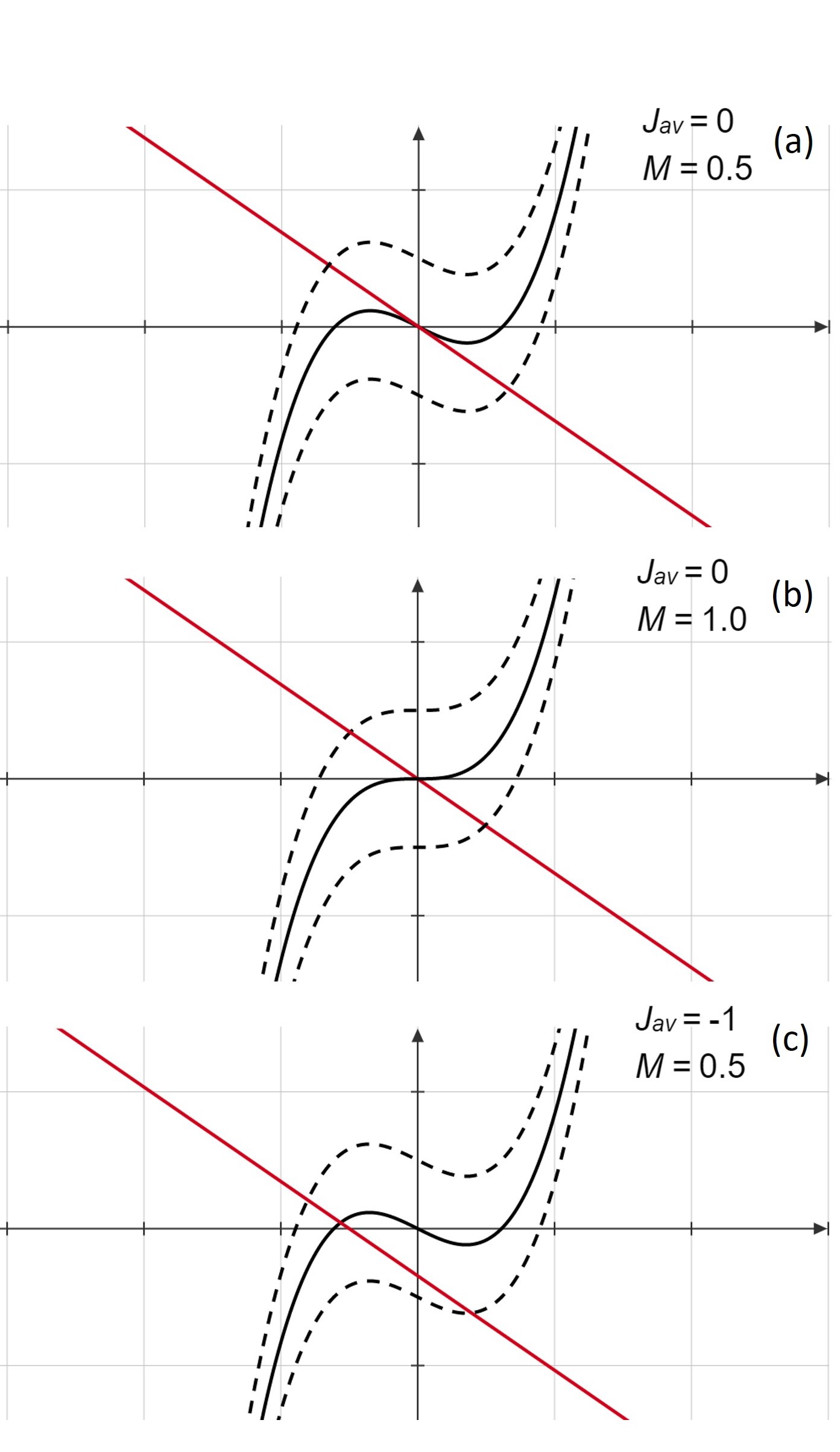}
\caption{Nullclines of Eqs.~(\ref{FNENG1}) and (\ref{FNENG2}) for different values of $\Jav$ and $M$. 
A comparison between panel (a) and (b) shows the effect of $M$ on the shape of the cubic nullcline. 
The area delimited by the dashed curves above and below the cubic nullcline in each panel illustrates the effect of instantaneous shifts caused by noise with an amplitude of up to $\pm 1$.}
\label{nullclines}
\end{figure}

It is instructive to observe the different impact of diversity and noise on the nullclines of Eqs.~(\ref{FNENG1}) and (\ref{FNENG2}). 
A change in diversity, i.e., in the standard deviation of the $J_i$ distribution, causes a change in $M$,
which affects the shape of the cubic nullcline by changing the coefficient of the linear term $X$ (see Fig.~\ref{nullclines}, panels (a) and (b)). 
This indicates that diversity can have a significant effect on overall network dynamics, independently of whether the mean value $\Jav$ is inside or outside the intrinsic oscillatory range $(-\varepsilon,+\varepsilon)$. 
On the other hand, the global noise term $\xi_G(t)$ can only cause rigid shifts, positive or negative, of the cubic nullcline along the vertical axis, as a consequence of its instantaneous fluctuations (compare the dashed and solid lines in Fig.~\ref{nullclines}).
This suggests that noise is unlikely to play a constructive role when diversity is optimized (i.e., in the conditions corresponding to a DIR) and $\Jav=0$. 
In this situation, noise will likely act as a perturbation of the system, which is already in an intrinsically oscillatory and resonant state.

Let us now consider what happens when $\Jav \neq 0$ (see Fig.~\ref{nullclines}-(c)).
In this case, the constant term $\Jav$ determines a rigid shift, upwards or downwards, of the second nullcline. 
This can significantly change the position of the equilibrium point of the system dynamics, turning the network from oscillatory into excitable. 
In these conditions, noise can play a synergistic role with diversity, by causing instantaneous rigid shifts of the cubic nullcline that counterbalance the effect of $\Jav$, thus triggering global network oscillations.

It should be noted that we did not add a periodic driving force to our system equations, of the type $A \sin(\Omega t)$. 
As mentioned in the introduction, this term is not necessary to observe either stochastic or diversity-induced resonance effects and its presence would introduce the additional constraint of matching two time scales, i.e., the driving force period and the oscillation period of the FitzHugh-Nagumo elements constituting the network.

\section{Numerical results and discussion}

In order to quantitatively study the combined effect of diversity and noise, we numerically solve the FitzHugh-Nagumo Eqs.~(\ref{FNEN1}) and (\ref{FNEN2}) for a network of $10^3$ elements with the above-mentioned topology and the following system parameters: $a=60$, $b=1.45$, $C=0.15$. 
The selected values of $a$ and $b$ generate a waveform and period similar to those of bursting oscillations of pancreatic $\beta$-cells~\cite{Scialla-2021a}. 
We set the coupling constant $C=0.15$, since, as we verified in an earlier work~\cite{Scialla-2021a}, the oscillatory response of the system is substantially unchanged by further increasing $C$ beyond $C=0.15$.

We run simulations for a range of diversity values $\sigma_d$ (from $\sigma_d =0$ to $\sigma_d =2.5$) and, at the same time, for a range of white noise standard deviation values $\sigma_n$ (from $\sigma_n =0$ to $\sigma_n =5.0$). 
We repeat this for each of the following diversity distribution mean values: $\Jav = 0, \pm 0.5, \pm 1$.

For each simulation, corresponding to a set of $\sigma_d$, $\sigma_n$, and $\Jav$ values, we quantify the network synchronization efficiency by computing the global oscillatory activity $\rho$~\cite{Cartwright-2000a,Scialla-2021a},
\begin{equation}
\rho = N^{-1} \sqrt{\left \langle [S(t) -\bar{S}]^2 \right \rangle} \, ,
\label{FNENG7}
\end{equation}
where $N = 10^3$ is the total number of oscillators, $S(t) = \sum_i x_i(t)$, and $\bar{S} = \langle S(t) \rangle$, with $\langle \dots \rangle$ denoting a time average. 
The results for the global oscillatory activity $\rho$ are plotted versus $\sigma_d$ and  $\sigma_n$, generating five three-dimensional surfaces corresponding to each of the above listed $\Jav$ values.

\begin{figure}[!t]
\includegraphics[width=8.0cm]{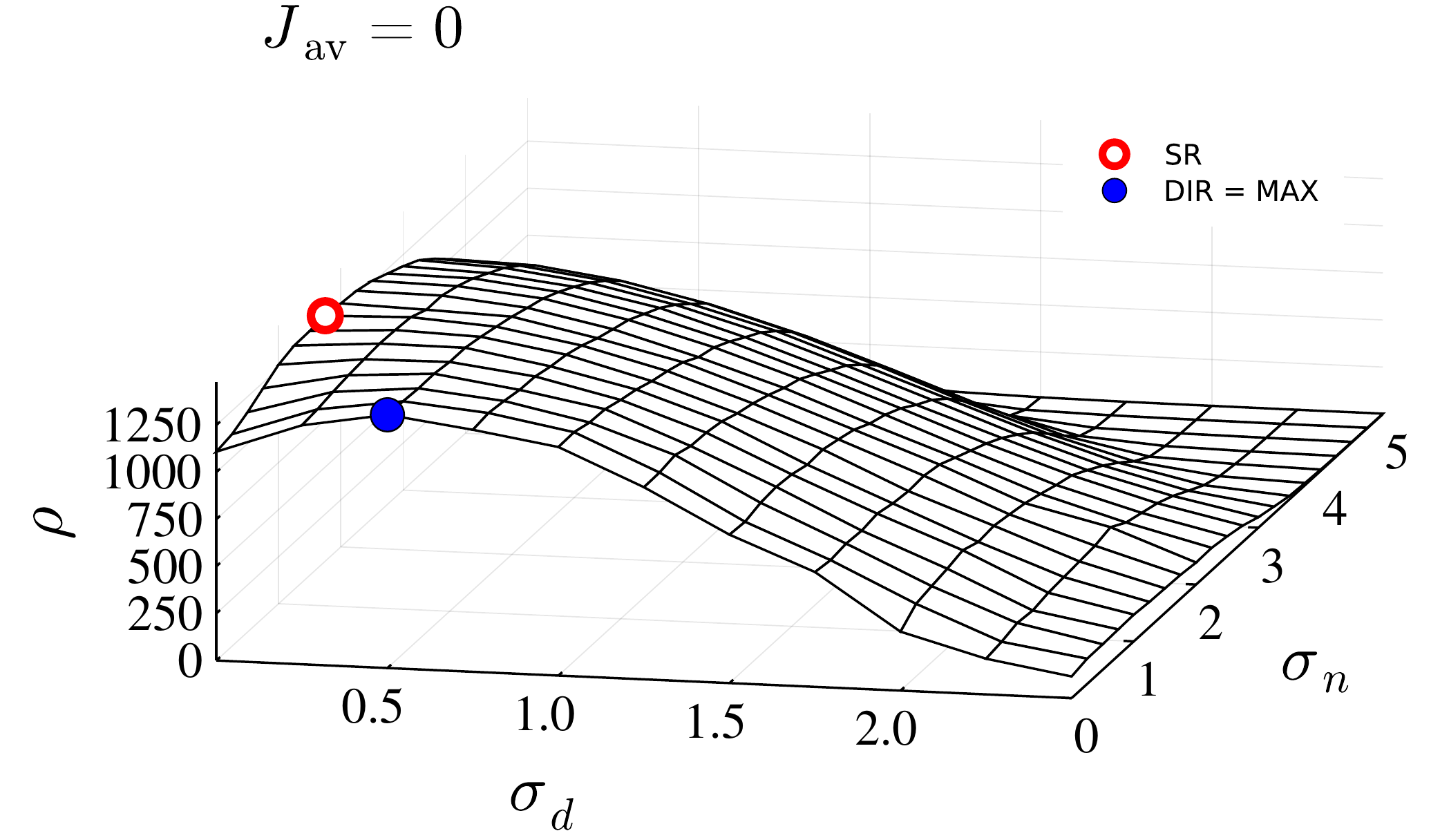}
\caption{Global oscillatory activity $\rho$, defined in Eq.~(\ref{FNENG7}), as a function of diversity ($\sigma_d$) and noise ($\sigma_n$), for $\Jav=0$. The full blue dot highlights the global surface maximum, which is coincident with the DIR maximum. The empty red dot corresponds to the noise-induced resonance maximum.}
\label{tevol-b0}
\end{figure}

In the $\Jav=0$ regime (Fig.~\ref{tevol-b0}), where a relatively high fraction or all of the network elements are inside the intrinsic oscillatory range, simulation results show that both diversity and noise are able to generate a resonance on their own. 
If we move along the diversity axis ($\sigma_n=0$, no noise) or along the noise axis ($\sigma_d=0$, no diversity), we observe in both cases a resonance maximum that is about 20-25\% higher than the $\rho$ value corresponding to the origin. 
In addition, the two sources of noise seem to act independently of one another, showing no evidence of a synergy. As a matter of fact, the global maximum of the surface coincides with the diversity-induced resonance maximum, therefore it occurs on the diversity axis, i.e., at $\sigma_d$=0.5, $\sigma_n$=0.
If, from this global maximum (shown as a full blue dot in Fig.~\ref{tevol-b0}), we move in any direction towards the middle of the surface, i.e., if we add noise, there is no gain in terms of collective oscillatory activity. This is consistent with the predictions of our qualitative theoretical analysis, indicating that noise is unlikely to play a constructive role for $\Jav=0$, when diversity is optimized.

\begin{figure}[!t]
\includegraphics[width=8.0cm]{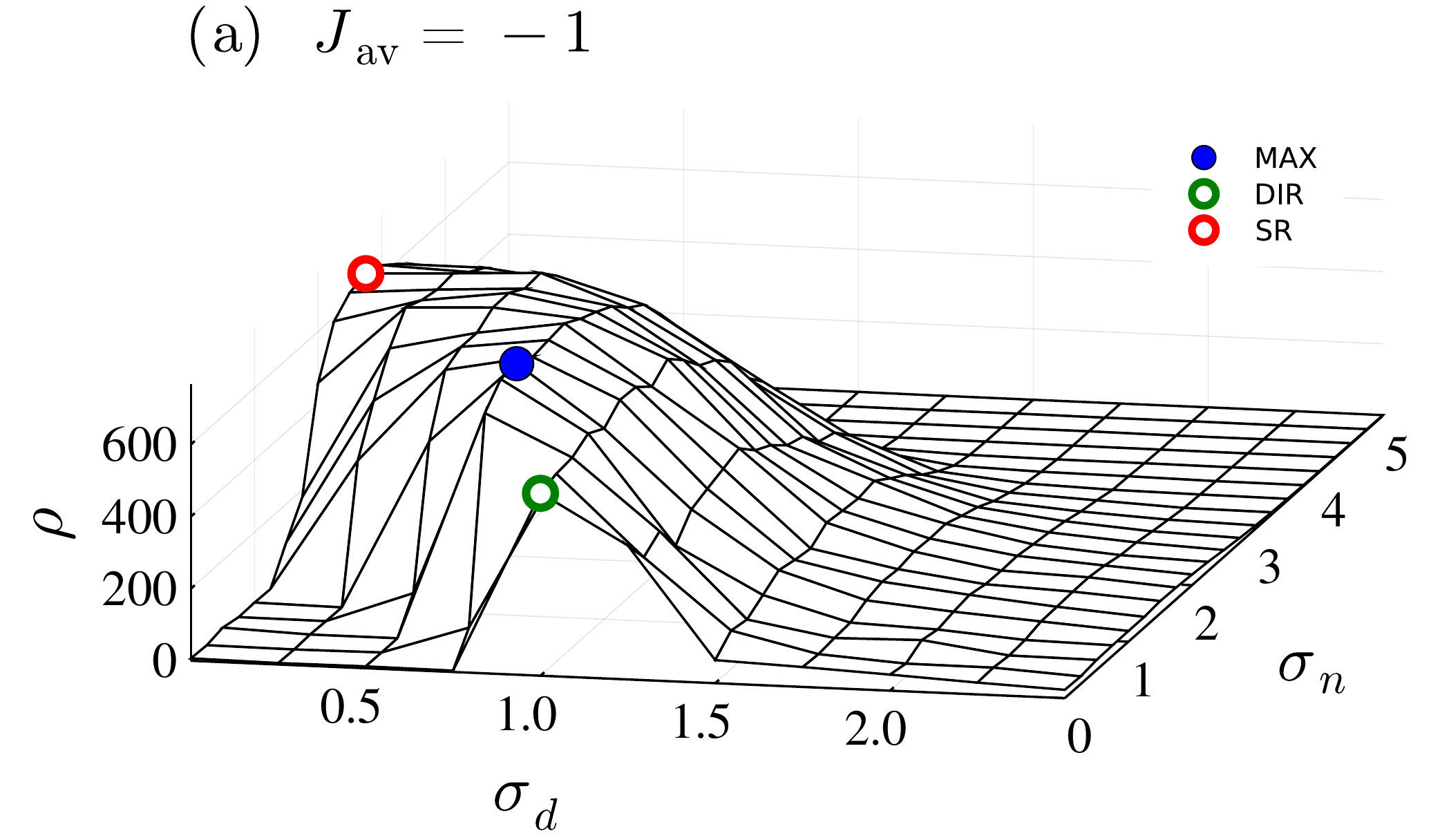}

\vspace{0.5cm}

\includegraphics[width=8.0cm]{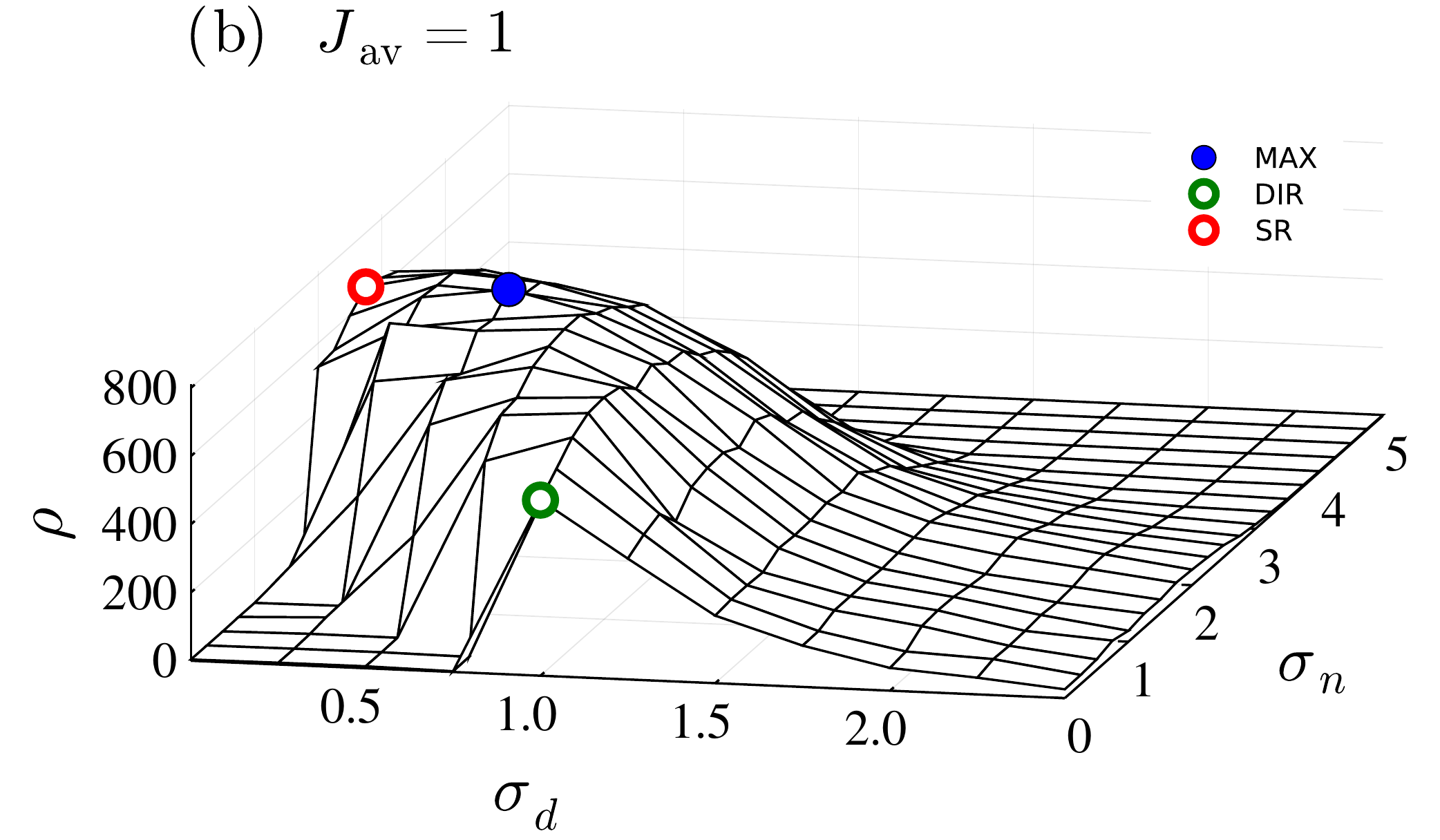}
\caption{Global oscillatory activity $\rho$, defined in Eq.~(\ref{FNENG7}), as a function of diversity ($\sigma_d$) and noise ($\sigma_n$), for $\Jav = -1$ (panel (a)) and $\Jav = +1$ (panel (b)). 
The full blue and empty green/red dots in each panel highlight the global surface maximum, the DIR maximum and the noise-induced resonance maximum, respectively.}
\label{fig-2}
\end{figure}

Moving to the opposite end of the $\Jav$ range, i.e., $\Jav = \pm1$ (Fig.~\ref{fig-2}), we observe a very different situation. In this regime most network elements are outside the intrinsic oscillatory range, either below ($\Jav =-1$) or above it ($\Jav =+1$). 
Here the addition of noise to diversity always results in a significant increase of the network oscillatory activity. 
In line with the theoretical analysis based on global system variables, this can be explained by considering that, in the case of $\Jav =-1$, most network elements are below the excitation threshold, i.e., in an excitable state, and can be pushed up into the oscillatory range by an instantaneous injection of positive external current, deriving from sufficiently large noise fluctuations with positive sign.
Vice versa, in the case of $\Jav =+1$, most elements are above the upper limit of the intrinsic oscillatory range, i.e., in an excitation block state, and can be pushed down into the oscillatory range by an instantaneous injection of negative external current, deriving from negative noise fluctuations with sufficiently large modulus. 
In both cases, the addition of noise on top of diversity causes a synergistic effect and a remarkable network synchronization improvement: for instance, the network oscillatory activity for $\Jav =+1$ raises by almost $50\%$, if we compare the DIR maximum ($\rho \approx 515$, empty green dot in Fig.~\ref{fig-2}, panel (b)), 
to the global maximum of the $\rho$ surface ($\rho \approx 738$, full blue dot in Fig.~\ref{fig-2}, panel (b)) resulting from the combination of diversity and noise effects.
It is also apparent from the data that, in this regime, noise is more efficient than diversity, as shown by the significantly higher noise-induced resonance maxima along the noise axis (empty red dots in Fig.~\ref{fig-2}, panel (a) and (b)), compared to their equivalents along the diversity axis (empty green dots in Fig.~\ref{fig-2}, panel (a) and (b)).

It is worth noting that the position of the DIR gets shifted towards higher values of $\sigma_d$ going from $\Jav = 0$ to $\Jav = \pm 1$. 
The DIR maximum is at $\sigma_d =0.5$ for $\Jav =0$, versus $\sigma_d =1$ for both $\Jav =-1$ and $\Jav =+1$ (empty green dots in Fig.~\ref{fig-2}, panels (a) and (b)). 
However, when we combine together noise and diversity, the position of the global maximum goes back to the same optimal diversity value found for $\Jav =0$
(full blue dots in Fig.~\ref{fig-2}, panels (a) and (b)). 
The mechanism of this effect is that noise stochastically ''throws'' network elements towards the oscillatory range, and it does so with respect to an average position on the $J$ axis that is determined, for each element, by its $J_i$ coefficient, deriving from the diversity distribution. When this mechanism reaches the highest efficiency, i.e., at the global maximum of the surface, the optimal diversity for $\Jav= \pm 1$ tends to be equal to that for $\Jav=0$. 
We may conclude that, in the $\Jav = \pm 1$ regime, there is a strong synergy between diversity and stochastic effects, which significantly broadens the range of resonant states of the network versus what can be observed when either source of disorder is applied individually.

\begin{figure}[!t]
\includegraphics[width=8.0cm]{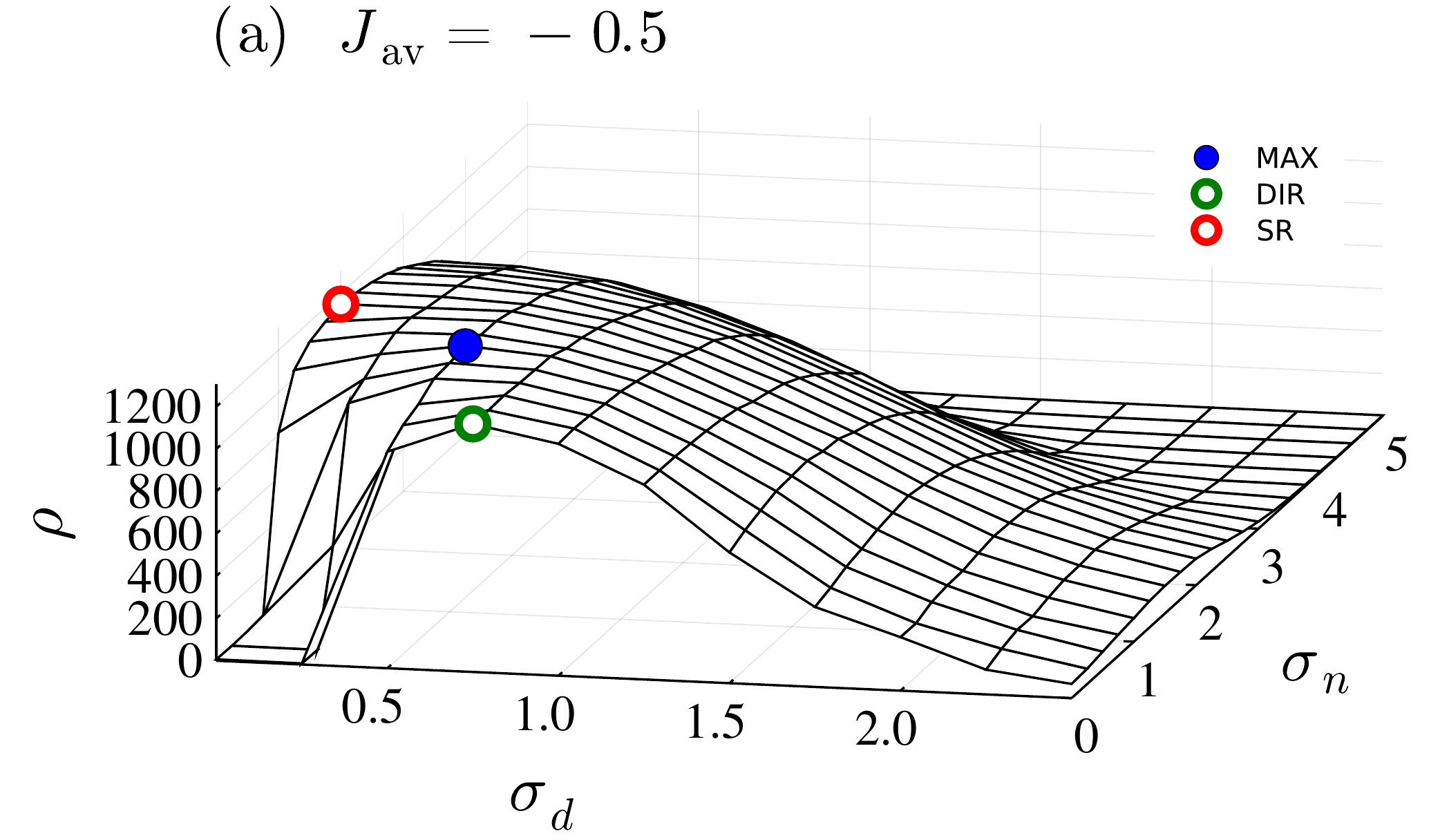}

\vspace{0.5cm}

\includegraphics[width=8.0cm]{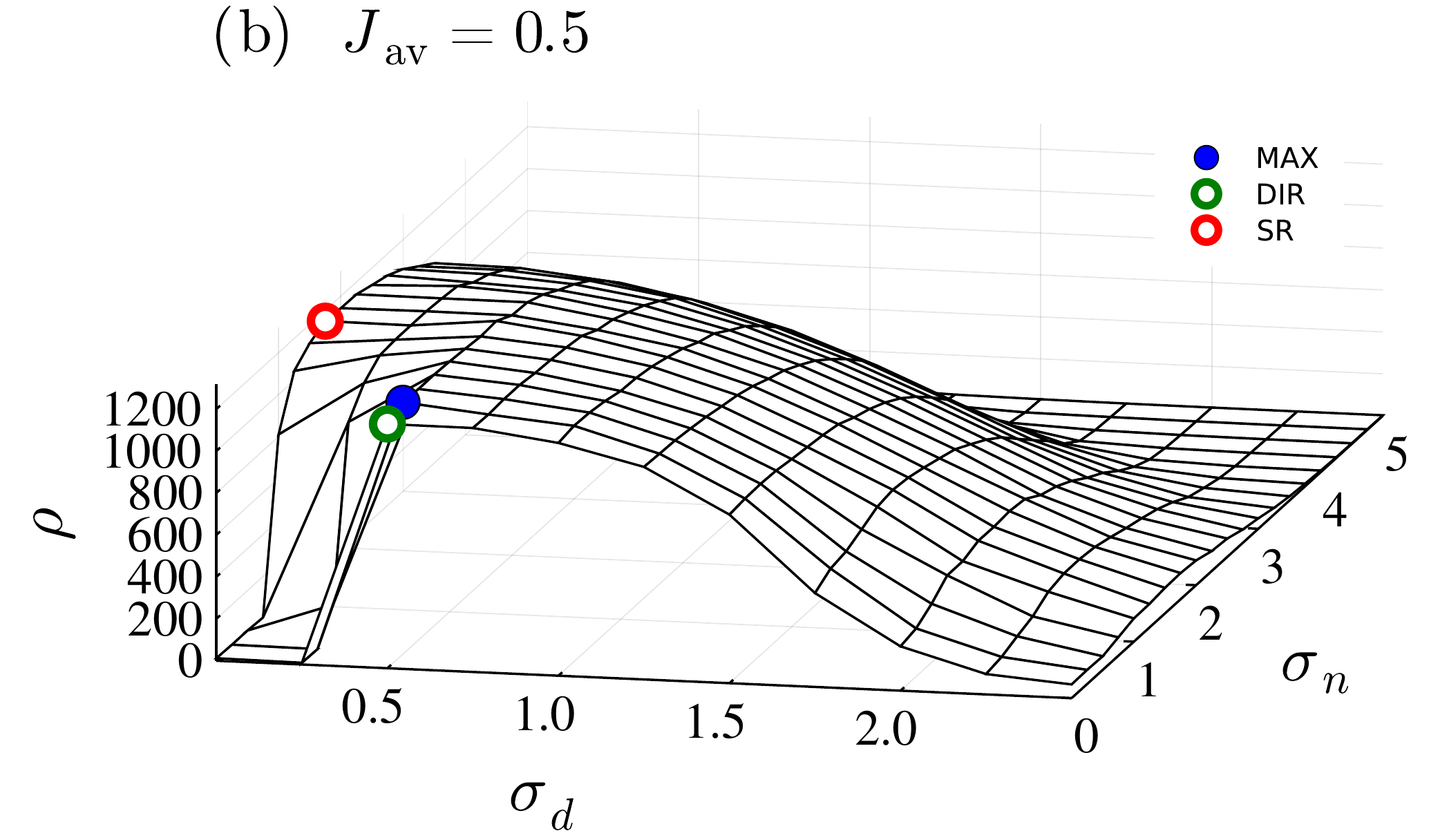}
\caption{Global oscillatory activity $\rho$, defined in Eq.~(\ref{FNENG7}), as a function of diversity ($\sigma_d$) and noise ($\sigma_n$), for $\Jav = -0.5$ (panel (a)) and $\Jav = +0.5$ (panel (b)). The full blue and empty green/red dots in each panel highlight the global surface maximum, the DIR maximum and the noise-induced resonance maximum, respectively.}
\label{fig-4}
\end{figure}

Finally, in the intermediate regime corresponding to $\Jav = \pm 0.5$ (Fig.~\ref{fig-4}), we observe an in between situation, with various regions of the $\rho$ surface where the combination of diversity and noise produces a synergy and an extension of the resonant range of the network. 
For example, at $\Jav =-0.5$, there are no network oscillations for diversity values $\sigma_d =0.0$ and $\sigma_d =0.25$, whereas, with the addition of noise, resonant states are observed in both cases, starting from $\sigma_n =1$ and $\sigma_n =0.5$, respectively. 
We point out that, also in this regime, the global maximum of the $\rho$ surface
due the combined diversity- and noise-induced resonance occurs, for $\Jav = -0.5$, at $\sigma_d =0.5$ (full blue dot in Fig.~\ref{fig-4}, panel (a)) and is shifted to smaller values with respect to the DIR maximum in the absence of noise ($\sigma_d =0.75$, empty green dot in Fig.~\ref{fig-4}, panel (a)). 
Therefore, the mechanism described in the previous paragraph, regarding the tendency of the optimal diversity value to be equal to that for $\Jav =0$, is at play here as well.

\begin{figure}[!t]
\includegraphics[width=8.0cm]{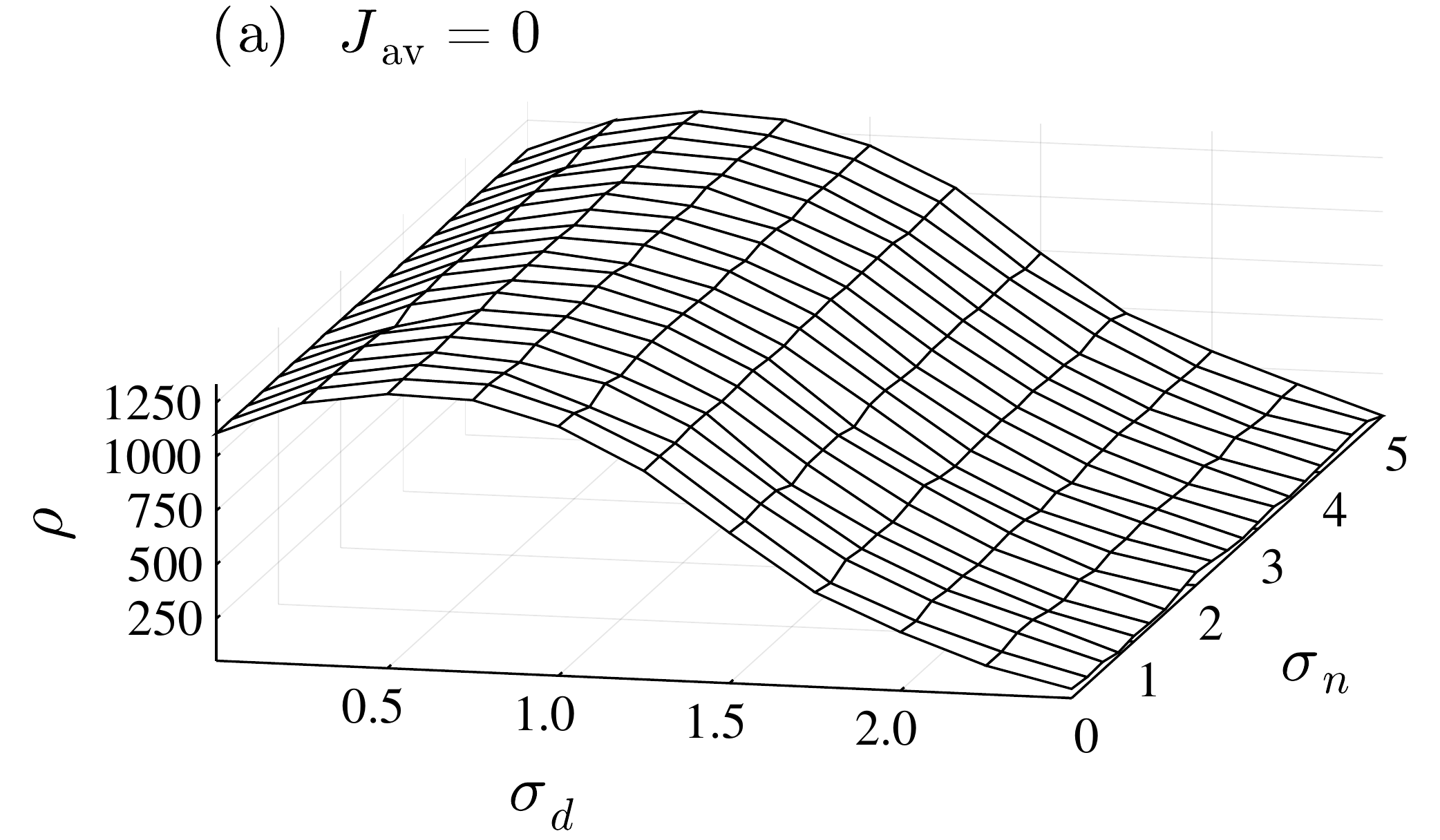}

\vspace{0.5cm}

\includegraphics[width=8.0cm]{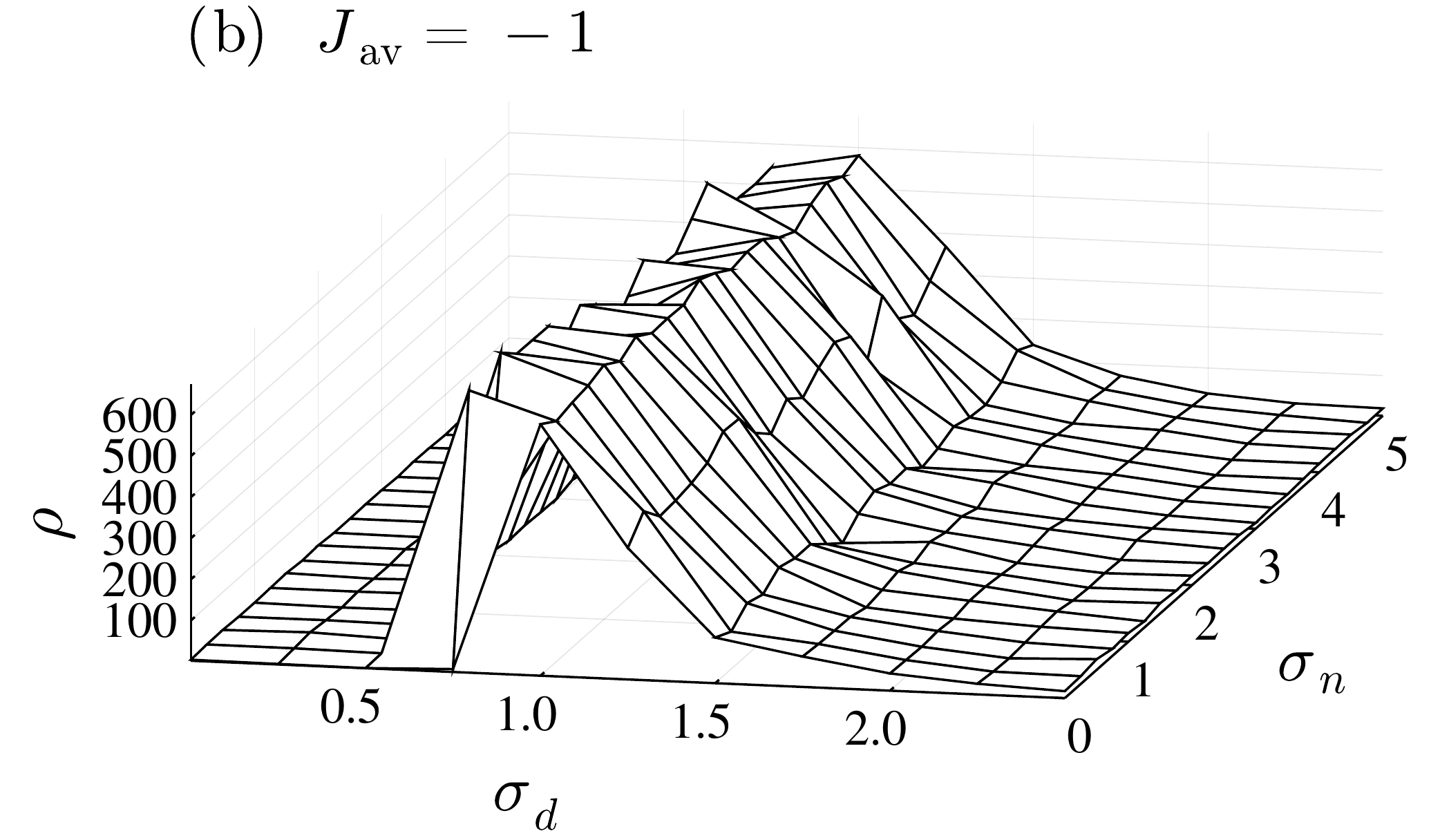}
\caption{Global oscillatory activity $\rho$, defined in Eq.~(\ref{FNENG7}), as a function of diversity ($\sigma_d$) and noise ($\sigma_n$), for $\Jav = 0$ (panel (a)) and $\Jav = -1$ (panel (b)), when noise is added into the second equation~(\ref{FNEN2}).}
\label{fig-7}
\end{figure}

In order to confirm the rationale for our choice of adding the noise term $\xi_i(t)$ into the first FitzHugh-Nagumo equation, Eq.~(\ref{FNEN1}), we also performed some simulations where $\xi_i(t)$ was added instead into the second equation, Eq.~(\ref{FNEN2}). 
We did this for $\Jav =0$ and $\Jav =-1$. 
As expected, the results reported in Fig.~\ref{fig-7} show that in this case the effect of noise is negligible and the network dynamics is entirely determined by diversity~\cite{degliesposti-2001a}.

%

\section{Conclusions}
Our theoretical and numerical analysis shows that, while there are some analogies between diversity- and noise-induced network synchronization, the two effects are substantially different and interact with each other differently, depending on the distance of the mean value of the diversity distribution from the intrinsic oscillatory range of the network elements. 
Specifically, when the diversity distribution is centered around the intrinsic oscillatory range ($\Jav =0$), diversity and noise act independently of one another and there is no indication of a synergy. On the other hand, when the mean value of the diversity distribution is far away from the intrinsic oscillatory range ($\Jav = \pm1$), then there is a clear synergy between the two sources of disorder, which determines a major improvement of network synchronization. In addition, in this regime, noise can improve network synchronization more effectively than diversity. This provides useful indications on the relative importance of the two effects in different network configurations, and on the possibility to neglect one or the other as a consequence.

Another important finding is that the optimal diversity value of the network is the same in all regimes, if noise is taken into account. 
In other words, when noise effects are added, the amount of diversity that maximizes collective oscillatory efficiency seems to be an intrinsic property of the network, independent of $\Jav$.

The fact that diversity and noise are not equivalent sources of disorder, but have distinct effects on network dynamics, may have implications for biological systems. 
Our results suggest that different network configurations can lead to a hierarchy between the two sources of disorder. This may have driven the exploitation of diversity and noise to a different degree in different biological systems during their evolution, depending on their specific nature and on the types of signals that trigger their activity.


\begin{acknowledgments}
    
The authors acknowledge support from the Estonian Research Council through Grant PRG1059
and the ERDF (European Development Research Fund) CoE (Center of Excellence) program through Grant TK133. 
The authors would like to thank Alessandro Loppini for helpful discussions.
	
\end{acknowledgments}



\end{document}